\documentclass[twocolumn,aps,pra]{revtex4-1}
\usepackage{epsfig}
\usepackage[english]{babel}
\usepackage{latexsym}
\usepackage{subfigure}
\usepackage{graphics}
\usepackage{epstopdf}
\usepackage{dcolumn}
\usepackage{amsmath}
\usepackage{hyperref}
\usepackage{amssymb}
\usepackage{appendix}
\usepackage{color}
\usepackage{lineno}
\usepackage{soul}
\usepackage{booktabs}
\usepackage{longtable}
\usepackage{multirow}
\usepackage{array}
\usepackage{ulem}
\usepackage{bm}

\begin{document}

\title{Isolated elliptically-polarized attosecond pulse generation in gapped graphene driven by linearly polarized laser fields}

\author{Xinru Song$^{1}$, Xiaoyu Bu$^{1}$, Xiaohui Zhao$^{1}$, Rongxiang Zhang$^{1}$, Shang Wang$^{2,\dag}$ and Fulong Dong$^{1,*}$,}

\date{\today}

\begin{abstract}
We theoretically investigate high-order harmonic generation (HHG) and its ellipticity in gapped graphene, driven by a femtosecond short-pulse laser at various orientation angles, employing the two-band density-matrix equations within the tight-binding approximation.
The orientation-dependent harmonic spectra exhibit pronounced enhancement of specific harmonics, which we attribute to the caustic effect.
Using the recombination trajectory model, we reveal that the orientation dependence of these enhanced harmonics originates from the distinct band structures encountered by electrons ionized from the two inequivalent $\textrm{K}$ points.
Moreover, we focus on the ellipticity of the enhanced harmonics at specific angles and demonstrate that it primarily depends on the phase difference between the parallel and perpendicular components, which can be accurately predicted by our recombination trajectory model.
Based on these insights, we propose a two-color (fundamental plus second-harmonic) field scheme to generate isolated elliptically polarized attosecond pulses (IEAPs) in gapped graphene.
Our findings may provide a promising pathway toward the generation of IEAPs in gapped graphene or transition metal dichalcogenides.

\end{abstract}
\affiliation{$^{1}$College of Physics Science and Technology, Hebei University, Baoding 071002, China\\
$^{2}$College of Physics and Hebei Key Laboratory of Photophysics Research and Application, Hebei Normal University, Shijiazhuang, 050024, China}

\maketitle

\section{Introduction}

Over the past decades, high-order harmonic generation (HHG) from atomic and molecular gases has been extensively studied \cite{Corkum1,ALHuillier,Lewenstein} due to its ability to produce attosecond pulses \cite{Ferenc}.
In recent years, much attention has been directed toward HHG in bulk crystals \cite{Luu,Ghimire} and two-dimensional (2D) materials \cite{CHeide,Keisuke}.
As a simple but special 2D material, the optical response of graphene under strong laser fields has been discussed in detail \cite{NYoshikawa,Dong1,Rost,Dong2}.
More recently, some attention has shifted to monolayer hexagonal boron nitride (hBN) \cite{LYue3,Dong6} and various transition metal dichalcogenides (TMDs) \cite{Hanzhe,NYoshikawa2}, which share similar lattice structures with graphene but exhibit broken spatial inversion symmetry.

Polarization is an intrinsic property of HHG \cite{FLanger}, and circularly-polarized pulses have long been desired for investigating the subfemtosecond dynamics of electrons in molecular systems \cite{Lischke,Handschin}.
The generation of isolated elliptically-polarized attosecond pulses (IEAPs) typically relies on sophisticated laser polarization control schemes \cite{KaiJYuan,LMedias,XiantuHe,WLiX}, inevitably increasing experimental complexity.
To address this question, high-ellipticity harmonics have been demonstrated under linearly polarized laser fields by tailoring the relative orientation between the laser and the target medium \cite{HDuLLuo,FLDong,Fulong,BMahieu}.
However, the realization of IEAPs remains a significant challenge.
This difficulty primarily arises from two factors:
1) Over a broad frequency range, the yield of perpendicular harmonics is generally much lower than that of the parallel component.
2) The relative phase between the two polarization components tends to be  unstable.

In monolayer hBN and TMDs, when the laser polarization deviates from the armchair direction, perpendicular anomalous harmonics, whose yields are comparable to those of the parallel harmonics, have been observed \cite{Hanzhe}.
The underlying mechanisms have been explored, and schemes for reconstructing the Berry curvature using these anomalous harmonics have also been proposed \cite{TTLuu,LYue2,AJUNar}.
Meanwhile, these experimental findings offer opportunities for synthesizing IEAPs.
However, several challenges remain:
1) The orientation dependence of the harmonic yield requires further theoretical investigation.
2) The stability of a relative phase close to $\pi/2$ between the parallel and perpendicular harmonics should be theoretically validated.
3) Interference effects among harmonics generated in different laser cycles need to be suppressed.

To address these challenges, we focus on gapped graphene with energy gaps of $0.05$ a.u. and $0.1$ a.u. (values comparable to those of TMDs such as $\textrm{MoS}_{2}$ and $\textrm{WS}_{2}$ \cite{KFMak,RLv}) and investigate its response to irradiation with a short-pulse laser.
We observe orientation-dependent enhanced harmonics, with relatively high ellipticity when the laser polarization is aligned along the zigzag direction.
By employing the electron-hole recombination trajectory model, we reveal that these enhanced harmonics arise from the caustic effect, and for specific trajectory branches, the phase difference between the parallel and perpendicular components can be locked at $\pi/2$.
Furthermore, we design a two-color (fundamental plus second-harmonic) field scheme to suppress interference among different harmonic channels, successfully generating isolated elliptically-polarized pulses on attosecond timescales.

This paper is organized as follows.
We describe our calculation methods for the two-band density-matrix equations (TBDMEs), and present the corresponding simulation results in Sec. \ref{s2}.
Section \ref{s3} presents our analytical model and reveals the underlying mechanisms of the enhanced harmonics and their ellipticity.
We design a two-color second-harmonic laser field scheme to generate elliptically-polarized attosecond pulses in Sec. \ref{s4}.
Finally, Sec. \ref{s5} presents our conclusion.
Throughout the paper, atomic units are used if not specified.

\section{Numerical calculation methods and results}
\label{s2}

\subsection{Numerical simulation methods}
\label{s2A}
The hexagonal lattice structure of gapped graphene is shown in Fig. 1(a), where $d= 1.42$ \AA \, denotes the carbon-carbon bond length.
Figure 1(b) illustrates the first Brillouin zone of gapped graphene, with the high-symmetry points $\Gamma$, $\textrm{M}$, $\textrm{K}$, and $\textrm{K}^{\prime}$.
In Figs. 1(a) and 1(b), the red arrows indicate the laser polarization direction, with $\theta$ defined as the angle between the polarization and the $x$-axis.

Within the tight-binding approximation and using Bloch states as basis vectors, the Hamiltonian derived from $\pi$ orbitals of gapped graphene can be expressed as
$
H_0=\left(\begin{array}{cc}
\Delta_g/2 & \gamma_0 f(\mathbf{k}) \\
\gamma_0 f^*(\mathbf{k}) & -\Delta_g/2
\end{array}\right).
$
Here $\gamma_0 = 0.1$ a.u. denotes the hopping energy, and $f(\textbf{k})=e^{i \texttt{k}_{x}d}+2\cos( \sqrt{3}\texttt{k}_{y}d / 2)e^{-i\texttt{k}_{x}d/2}$.
The energy eigenvalues of the conduction $(c)$ and valence $(v)$ bands are $\varepsilon_c(\mathbf{k})=-\varepsilon_v(\mathbf{k})= \sqrt{\gamma_0^2 |f(\mathbf{k})|^2 + \Delta_g^{2}/4}$, as illustrated in Fig. 1(c).

\begin{figure}[t]
\begin{center}
\includegraphics[width=7cm,height=7cm]{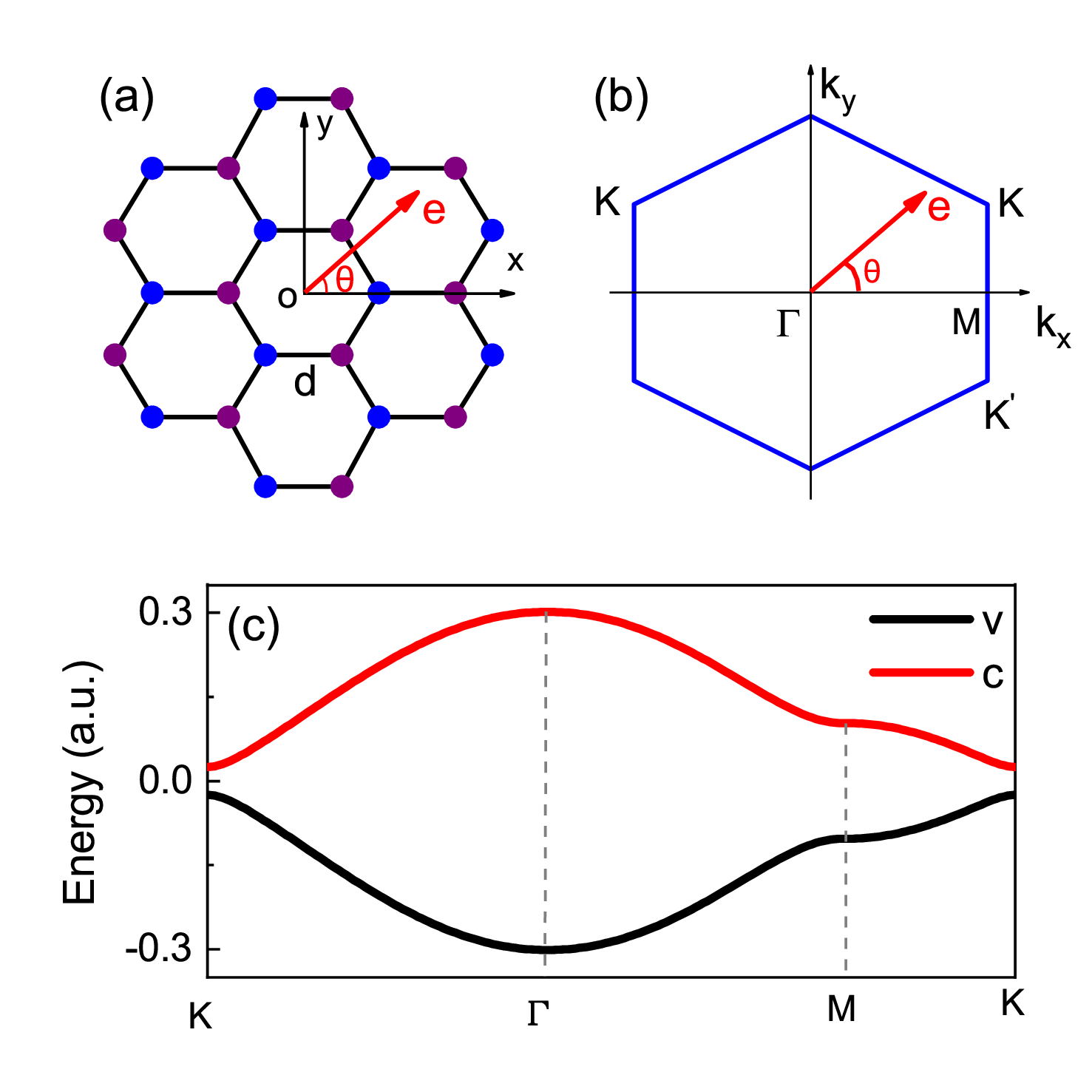}
\caption{
(a) Hexagonal lattice structure of gapped graphene.
(b) First Brillouin zone of the corresponding reciprocal lattice.
The angle $\theta$ denotes the orientation between laser polarization direction $\textbf{\textit{e}}$ and the $x$-axis.
(c) Electronic band structure of the $v$ and $c$ bands along the $\textrm{K}$-$\Gamma$-$\textrm{M}$-$\textrm{K}$ path in the Brillouin zone for gapped graphene.
}
\label{fig:graph1}
\end{center}
\end{figure}

We numerically simulate the currents in gapped graphene by solving TBDMEs in the Houston representation \cite{WVHouston}.
Within the dipole approximation, these equations are

\vspace{-0.4cm}
\begin{align}
i \frac{d}{d t} \rho_{m n}^{\mathbf{K}_{t}}(t) &= \left[\varepsilon_{mn} (\mathbf{K}_{t}) - i \tilde{\delta}_{mn}/T_2 \right] \rho_{m n}^{\mathbf{K}_{t}}(t)  \nonumber\\
&+ \textit{\textbf{F}}(t) \cdot \sum_{l} \left[\mathbf{D}_{m l}^{\mathbf{K}_{t}} \rho_{l n}^{\mathbf{K}_{t}}(t)-\mathbf{D}_{l n}^{\mathbf{K}_{t}} \rho_{m l}^{\mathbf{K}_{t}}(t)\right],
\end{align}
where $\hat{\rho}$ is the density matrix, with elements $\rho_{mn}^{\mathbf{k}}=\langle \phi_m^{\mathbf{k}} (\mathbf{r}) \vert \hat{\rho} \vert \phi_n^{\mathbf{k}} (\mathbf{r}) \rangle$.
Here, $m$ and $n$ denote $v$ or $c$ bands.
$\varepsilon_{mn} (\mathbf{k}) = \varepsilon_{m} (\mathbf{k}) - \varepsilon_{n} (\mathbf{k})$ represents the energy difference between bands $m$ and $n$.
$\tilde{\delta}_{mn}=1-\delta_{mn}$, and the dephasing rate is $1/T_2=0.01$ a.u.
The transition dipole elements are given by $\textrm{\textbf{D}}_{mn}^{\textbf{k}} = i \langle u_{m,\textbf{k}}(\textbf{r}) \vert \nabla_{\textbf{k}} \vert u_{n,\textbf{k}}(\textbf{r}) \rangle$, where $u_{m,\textbf{k}}(\textbf{r})$ denotes the periodic part of the Bloch wavefunction for $m$ band of gapped graphene \cite{GVampa,SCJiang}.

In Eq. (1), the crystal quasimomentum is given by $\textbf{K}_{t} = \textbf{K}_{0} + \textit{\textbf{A}}(t)$, where $\mathbf{K}_0$ lies within the first Brillouin zone.
The vector potential $\textit{\textbf{A}}(t)$ is defined as  $\textit{\textbf{A}}(t) = - \int^{t} \textit{\textbf{F}}(\tau) d\tau$, where $\textit{\textbf{F}}(t) \equiv F(t) \textit{\textbf{e}} = F_0 f(t) \cos \left(\omega_0 t\right) \textit{\textbf{e}} $ is the electric field of the laser pulse.
Here, $f(t) = \sin^{2}(\omega_0 t/2n)$ is the laser envelope function with $n=3$.
The frequency $\omega_0$ corresponds to a wavelength of $4000$ nm,
and the amplitude $F_0$ corresponds to an intensity of $1 \times 10^{12}$ W/cm$^2$.
The unit vector $\textit{\textbf{e}}$ indicates the polarization direction of the electric field, as shown in Figs. 1(a) and 1(b).

Equation (1) can be readily solved numerically using the standard fourth-order Runge-Kutta algorithm.
The interband, intraband and total currents are evaluated as

\vspace{-0.4cm}
\begin{subequations}
\begin{align}
\textbf{\textit{j}}_{\text {inter}}(t) & = \int_{\textrm{BZ}} d \mathbf{K}_{0} \sum_{m \neq n} \rho_{n m}^{\mathbf{K}_{t}}(t) \mathbf{P}_{m n}^{\mathbf{K}_{t}}, \\
\textbf{\textit{j}}_{\mathrm{intra}}(t) & = \int_{\textrm{BZ}} d \mathbf{K}_{0} \sum_n \rho_{n n}^{\mathbf{K}_{t}}(t) \mathbf{P}_{n n}^{\mathbf{K}_{t}}, \\ \textbf{\textit{j}}_{\textrm{tot}}(t) & =\textbf{\textit{j}}_{\text {inter}}(t)+\textbf{\textit{j}}_{\mathrm{intra}}(t).
\end{align}
\end{subequations}
Here, $\mathbf{P}_{cc}^{\mathbf{k}} = \nabla_{\mathbf{k}} \varepsilon_{c}(\mathbf{k}) = -\mathbf{P}_{vv}^{\mathbf{k}}$ denotes the group velocity, and $\mathbf{P}_{cv}^{\mathbf{k}} = i (\varepsilon_{c}(\mathbf{k}) - \varepsilon_{v}(\mathbf{k})) \mathbf{D}_{cv}(\mathbf{k})$ represents the interband current matrix element.
The total harmonic yield is calculated by
\begin{eqnarray}
\begin{aligned}
H_{\textrm{tot}}(\omega)=\omega^{2} (\vert j_{\parallel}(\omega) \vert^{2} + \vert j_{\perp}(\omega) \vert^{2}),
\end{aligned}
\end{eqnarray}
where $\textit{j}_{\mu}(\omega)= \int^{\infty}_{-\infty} \textit{\textbf{e}}_{\mu} \cdot \textbf{\textit{j}}_{\textrm{tot}}(t) e^{-i \omega t} dt$,
and $\mu$ denotes $\parallel$ or $\perp$.

\begin{figure}[t]
\begin{center}
\includegraphics[width=8.5cm,height=7cm]{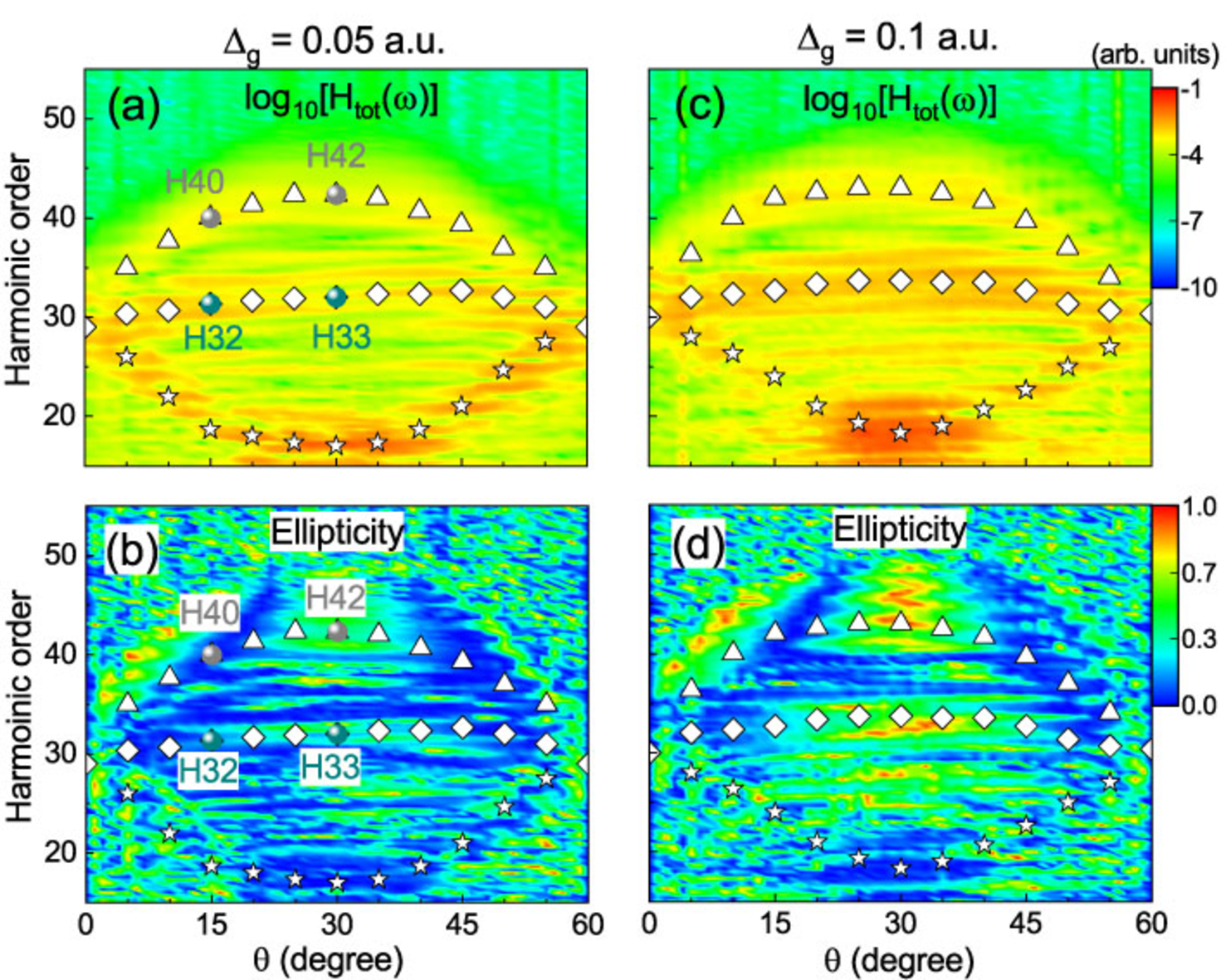}
\caption{
(a) Total harmonic intensity as a function of the orientation angle, calculated using Eq. (3), for gapped graphene with $\Delta_{g} = 0.05$ a.u.
(b) Harmonic ellipticity corresponding to the HHG in (a), calculated by Eq. (4).
(c), (d) Same as (a) and (b), respectively, but for $\Delta_{g} = 0.1$ a.u.
In panels (a) and (c), the small triangles, rhomboids, and stars indicate the prominent peaks in the harmonic intensity, and these markers are also translated to panels (b) and (d) for reference.
}
\label{fig:graph1}
\end{center}
\end{figure}

The harmonic ellipticity can be evaluated using the following relation \cite{zhushiyi}:
\begin{eqnarray}
\begin{aligned}
\varepsilon = \sqrt{\dfrac{1+r^{2}-\sqrt{1+2r^{2}\cos 2\delta + r^{4}}}{1+r^{2}+\sqrt{1+2r^{2} \cos 2\delta + r^{4}}}},
\end{aligned}
\end{eqnarray}
where $r(\omega)=\vert j_{\perp}(\omega) \vert / \vert j_{\parallel}(\omega)\vert$ is the amplitude ratio between the perpendicular and parallel components.
The phase difference is given as $\delta(\omega) =\vert \varphi_{\parallel}(\omega) - \varphi_{\perp}(\omega) + q \pi \vert$, constrained to the range of $[0,\pi/2]$, where $\varphi_{\mu}(\omega) =$ arg$[j_{\mu}(\omega)]$ and $q$ is an integer.
From Eq. (4), it can be concluded that high ellipticity is expected when the amplitude of the perpendicular harmonic component is comparable to that of the parallel component, and the phase difference $\delta$ approaches $\pi/2$.

The time-frequency distribution of the total current can be evaluated by
\begin{align}
H^{\textrm{tf}}(\omega,t) = \omega^2 (\vert \textit{j}^{\textrm{tf}}_{\parallel}(\omega,t) \vert^2 + \vert \textit{j}^{\textrm{tf}}_{\perp}(\omega,t) \vert^2),
\label{eq:4}
\end{align}
in which $\textit{j}^{\textrm{tf}}_{\mu}(\omega,t) = \int_{t-T}^{t+T} d t^{\prime} \textit{\textbf{e}}_{\mu} \cdot \textbf{\textit{j}}_{\textrm{tot}}(t^{\prime}) W(t^{\prime} - t) e^{i \omega t^{\prime}}$, here $W(x) = \dfrac{1}{\sqrt{2 \pi \tau_0}} e^{-x^2 / 2\tau_0^2}$ with $\tau_0 = 1/3 \omega_0$.
$T = 2 \pi / \omega_0$ is the laser period.

\subsection{Numerical simulation results}
\label{s2B}

Figure 2(a) illustrates the orientation dependence of the total harmonic yield for gapped graphene with $\Delta_g = 0.05$ a.u., calculated by Eq. (3).
At $\theta = 0^{\circ}$, a pronounced harmonic enhancement appears around the $29^{\textrm{th}}$ harmonic (H29), similar to what has been observed in graphene \cite{Dong5}.
As the orientation angle increases, this prominent enhancement splits into three distinct branches, which are marked by small triangles, rhomboids, and stars, respectively.
When $\theta$ reaches $60^{\circ}$, these three branches merge again near H29.

Figure 2(b) presents harmonic ellipticity corresponding to the spectra shown in Fig. 2(a).
For the enhanced harmonics indicated by small triangles and rhomboids, the ellipticity exhibits a decreasing trend as the orientation angle increases from $0^{\circ}$ to $15^{\circ}$, followed by a gradual increase as the angle further increases from $15^{\circ}$ to $30^{\circ}$.
Specifically, at $\theta = 15^{\circ}$, the ellipticity of the enhanced $32^{\textrm{nd}}$ and $40^{\textrm{th}}$ harmonics (H32 and H40) is relatively low (notably, the ellipticity of H40 is nearly zero).
In contrast, at $\theta = 30^{\circ}$, the ellipticity of the $33^{\textrm{rd}}$ and $42^{\textrm{nd}}$ harmonics (H33 and H42) is relatively high.
It is worth noting that we do not primarily focus on the ellipticity of the harmonics marked by the small stars, as they exhibit relatively disordered behavior.

Similarly, Figs. 2(c) and 2(d) show the harmonic spectra and ellipticity, respectively, for gapped graphene with $\Delta_g = 0.1$ a.u.
The simulation results in Figs. 2(c) and 2(d) are qualitatively consistent with those in Figs. 2(a) and 2(b), although the intensity of enhanced harmonics is stronger and the corresponding ellipticity is higher.
In the following, we use the simulation results of $\Delta_g = 0.05$ a.u. to analyze the underlying mechanisms of the orientation dependence of the harmonic enhancement structure and ellipticity.

\section{Analytical model and mechanism discussion}
\label{s3}
\subsection{Electron-hole recombination trajectory model}
\label{s3A}

To elucidate the underlying mechanism, we investigate the recombination trajectory model of electron-hole pairs within the framework of the two-band density matrix equation.
Under the strong-field approximation, intraband currents can be neglected, while the interband current plays a dominant role in the process of HHG.
In this approximation, the Fourier transform of the interband current is

\vspace{-0.4cm}
\begin{align}
\textit{j}_{\mu}(\omega) \sim & \int d \textbf{K}_{0} \int_{- \infty}^{\infty} d t \int_{-\infty}^{t} dt^{\prime} g_{\mu}(\textbf{K}_{0},t^{\prime},t) e^{-iS_{\mu}(\textbf{K}_{0},t^{\prime},t,\omega)},
\end{align}
in which
$g_{\mu}(\textbf{K}_{0},t^{\prime},t) = -\textit{F}(t^{\prime}) \left| \textrm{D}_{cv,\parallel}^{\mathbf{K}_{t^{\prime}}} \right|  \varepsilon_{c v} (\textbf{K}_{t}) \left| \textrm{D}_{cv,\mu}^{\mathbf{K}_{t}} \right|$ is a slowly varying term.
In Eq. (6), the semiclassical action is

\vspace{-0.4cm}
\begin{align}
 S_{\mu} (\textbf{K}_{0},t^{\prime},t,\omega) = & \int_{t^{\prime}}^t  \left[ \varepsilon_{c v} (\textbf{K}_{\tau}) + \boldsymbol{F}(\tau) \cdot \boldsymbol{\mathcal{A}}(\textbf{K}_{\tau}) \right] d \tau \nonumber \\
& + \alpha_{\mu}(\textbf{K}_{t}) - \alpha_{\parallel}(\textbf{K}_{t^{\prime}})  - \omega t,
\end{align}
in which $\boldsymbol{\mathcal{A}}(\textbf{k}) = \textbf{D}_{c c}^{\textbf{k}} -\textbf{D}_{v v}^{\textbf{k}}$, and
$\alpha_{\mu}(\textbf{k}) = \textrm{arg}(\textrm{D}_{c v,\mu}^{\textbf{k}})$ is the transition dipole phase.
Here, $\textrm{D}_{c v,\parallel}^{\textbf{k}} = \textrm{D}_{c v,x}^{\textbf{k}} \cos \theta + \textrm{D}_{c v,y}^{\textbf{k}} \sin \theta$ and $\textrm{D}_{c v,\perp}^{\textbf{k}} = -\textrm{D}_{c v,x}^{\textbf{k}} \sin \theta + \textrm{D}_{c v,y}^{\textbf{k}} \cos \theta$, in which $\textrm{D}_{c v,x(y)}^{\textbf{k}}$ denotes the $x$ ($y$) component of $\textbf{D}_{c v}^{\textbf{k}}$.

Next, we apply the steady-phase approximation to the variables $\textbf{\textrm{K}}_0, t, t^{\prime}$, yielding the following equations:

\vspace{-0.4cm}
\begin{subequations}
\begin{align}
\varepsilon_{c v}(\mathbf{K}^{\textrm{st}}_{t_{i}}) + & \boldsymbol{F}(t_{i}) \cdot \left( \boldsymbol{\mathcal{A}}(\textbf{K}^{\textrm{st}}_{t_{i}})-\nabla_{\mathbf{K}^{\textrm{st}}_{t_{i}}} \alpha_{\|}(\mathbf{K}^{\textrm{st}}_{t_{i}}) \right) \leq \varepsilon_{i}, \\
\bigg|  \int_{t_i}^{t_r} \nabla_{\mathbf{K}^{\textrm{st}}_{\tau}} & \left( \varepsilon_{c v}(\mathbf{K}^{\textrm{st}}_{\tau}) + \boldsymbol{F}(\tau) \cdot \boldsymbol{\mathcal{A}}(\mathbf{K}^{\textrm{st}}_{\tau}) \right)   d \tau \nonumber \\
+ & \nabla_{\mathbf{K}^{\textrm{st}}_{t_r}}  \alpha_{\mu}(\mathbf{K}^{\textrm{st}}_{t_r}) - \nabla_{\mathbf{K}^{\textrm{st}}_{t_i}} \alpha_{\parallel}(\mathbf{K}^{\textrm{st}}_{t_i}) \bigg|  =0 , \\
\varepsilon_{c v}(\mathbf{K}^{\textrm{st}}_{t_r}) + & \boldsymbol{F}(t_r) \cdot\left(\boldsymbol{\mathcal{A}}(\mathbf{K}^{\textrm{st}}_{t_r})-\nabla_{\mathbf{K}^{\textrm{st}}_{t_r}} \alpha_{\mu}(\mathbf{K}^{\textrm{st}}_{t_r})\right)=\omega.
\end{align}
\end{subequations}
In Eq. (8), $t_i$ and $t_r$ represent the ionization and recombination times of electron-hole pairs, respectively.
Here, $\mathbf{K}^{\textrm{st}}_t = \mathbf{K}_0^{\textrm{st}} + \textit{\textbf{A}}(t)$, where $\mathbf{K}_0^{\textrm{st}} = (\textrm{K}_{0x}^{\textrm{st}},\textrm{K}_{0y}^{\textrm{st}})$ is the saddle-point momentum.

In the process of solving the equation, we first sample the lattice momenta $\textbf{\textrm{K}}_0$ from the first Brillouin zone, which correspond to inequivalent electrons.
Next, the electrons oscillate in the reciprocal space driven by the laser.
At time $t_{i}$, determined by the condition in Eq. (8a), electrons are excited from the $v$ to $c$ band, leaving a hole in the $v$ band.
It is important to note that electron excitation can occur not only at the point of minimum energy gap (i.e., the $\textrm{K}$ point), but also in its vicinity \cite{Dong5,MKolesik,LunYue}.
In our simulation, we set the excitation energy threshold $\varepsilon_{i}$ to $0.1$ a.u. for gapped graphene of $\Delta_g = 0.05$ a.u.
Next, the electron-hole pairs are assumed to move in the two-dimensional coordinate space.
When they recombine perfectly, which implies Eq. (8b) is satisfied, the recombination time $t_r$ is obtained.
Finally, the collision energy $\omega$ can be calculated using Eq. (8c).

Under the steady-phase approximation, for a specific saddle-point momentum $\textbf{K}_{0}^{\textrm{st}}$, Eq. (6) can be integrated to
$\textit{j}_{\mu}^{\textbf{K}_{0}^{\textrm{st}}}(\omega) \varpropto  g_{\mu}(\textbf{K}_{0}^{\textrm{st}},t_i,t_r) \dfrac{e^{-iS_{\mu}(\textbf{K}_{0}^{\textrm{st}},t_i,t_r,\omega)}}{\sqrt{\vert \operatorname{det} [S_{\mu}^{\prime \prime}(\textbf{K}_{0}^{\textrm{st}},t_i,t_r,\omega)] \vert}}$,
where $S_{\mu}^{\prime \prime}(\textbf{K}_{0}^{\textrm{st}},t_i,t_r,\omega)$ is the Hessian matrix \cite{JChen,Dong5}.
Therefore, for a specific saddle-point trajectory, the phase difference $\delta$ between $\textit{j}_{\parallel}^{\textbf{K}_{0}^{\textrm{st}}}(\omega)$ and $\textit{j}_{\perp}^{\textbf{K}_{0}^{\textrm{st}}}(\omega)$ can be evaluated by

\vspace{-0.4cm}
\begin{align}
\delta (\textbf{K}_{0}^{\textrm{st}}) = S_{\parallel}(\textbf{K}_{0}^{\textrm{st}}) - S_{\perp}(\textbf{K}_{0}^{\textrm{st}}) = \alpha_{\parallel}(\mathbf{K}^{\textrm{st}}_{t_r})-\alpha_{\perp}(\mathbf{K}^{\textrm{st}}_{t_r}).
\end{align}

\subsection{Orientation dependence of the harmonic enhancement structure}
\label{s3B}

Figure 3(a) presents the time-frequency distribution of the total harmonic spectrum at $\theta = 0^\circ$, calculated using Eq. (5).
The recombination energy $\omega(t_r)$ and the saddle-point momenta $\textbf{\textrm{K}}_{0}^{\textrm{st}}$, obtained from Eq. (8), are overlaid in Figs. 3(a) and 3(b), respectively.
The results in Fig. 3(a) demonstrate that the saddle-point solution $\omega(t_r)$ qualitatively reproduces the numerical time-frequency distribution.
Figure 3(b) reveals that within one laser cycle, four saddle-point momentum branches, labeled by $\textrm{B}_1$-$\textrm{B}_4$ in the first Brillouin zone, dominate the HHG process.
Corresponding electrons are excited into $c$ band in the vicinity of $\textrm{K}$ points indicated by the gray dashed arrows.
Due to the symmetry of gapped graphene at $\theta = 0^{\circ}$, the recombination trajectories $\omega(t_r)$ between $\textrm{B}_1$ and $\textrm{B}_3$, and between $\textrm{B}_2$ and $\textrm{B}_4$, overlap in Fig. 3(a).

Previous studies have shown that when the combined trajectories satisfy $d \omega / d t_i =0$, (that is, when the long and short orbits converge) significant harmonic enhancement occurs, a phenomenon known as the caustic effect \cite{ORaz,DFacciala}.
In Fig. 3(a), it is evident that for the branches $\textrm{B}_1$ and $\textrm{B}_3$, the convergence of the long and short orbits occurs near H30, while for $\textrm{B}_2$ and $\textrm{B}_4$, it occurs around H26.
The time-frequency distribution shows a clear enhancement of harmonic yields at these caustic points, consistent with the numerical results at $\theta = 0^{\circ}$ in Fig. 2(a), where significant enhancement is observed between H26 and H30.

\begin{figure}[t]
\begin{center}
\includegraphics[width=8.5cm,height=7cm]{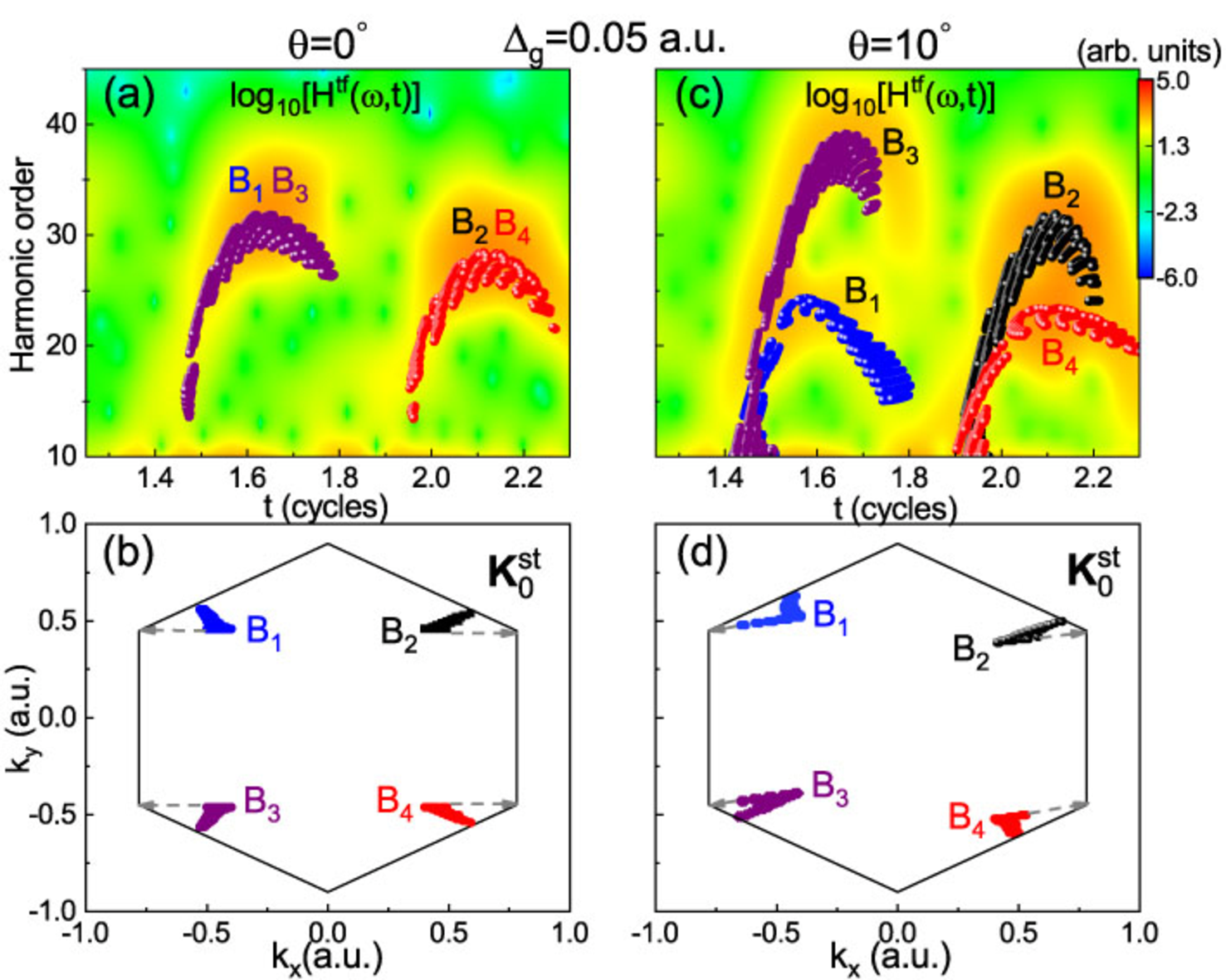}
\caption{(a) Time-frequency distribution of the total current for an orientation angle of $\theta = 0^{\circ}$, calculated using Eq. (5).
The overlaid points represent the recombination energies as a function of the recombination time, $\omega (t_r)$, obtained from Eq. (8).
(b) Saddle-point momenta $\textbf{\textrm{K}}_{0}^{\textrm{st}}$ corresponding to the electron trajectories in (a), calculated using Eq. (8).
(c), (d) Same as (a) and (b), respectively, but for an orientation angle of $\theta = 10^\circ$.
}
\label{fig:graph1}
\end{center}
\end{figure}

When the orientation angle increases to $10^{\circ}$, the situation is significantly different.
At this angle, four distinct orbital branches within a single laser cycle can be clearly identified in both saddle-point momenta (Fig. 3(d)) and the time-frequency distribution (Fig. 3(c)).
As shown in Fig. 3(c), all four branches clearly exhibit long and short orbits.
For branches $\textrm{B}_1$ and $\textrm{B}_4$, their convergence occurs near H22.
For $\textrm{B}_2$, the caustic effect appears at H30, and for $\textrm{B}_3$, at H37.
The corresponding enhancements in the time-frequency distribution align well with the harmonics marked by stars, rhomboids, and triangles at $10^{\circ}$ in Fig. 2(a), respectively.

In Fig. 3(c), clear differences can be observed between branches $\textrm{B}_1$ and $\textrm{B}_3$, as well as between $\textrm{B}_2$ and $\textrm{B}_4$.
Figure 3(d) reveals that these differences arise because, at an orientation angle of $10^{\circ}$, the saddle-point electrons will be excited from two inequivalent $\textrm{K}$ points to $c$ band.
The resulting the electron-hole pairs then experience different band structures, leading to different coordinate space motions and recombination energies.
In addition, the differences between $\textrm{B}_1$ and $\textrm{B}_4$, as well as between $\textrm{B}_2$ and $\textrm{B}_3$, originate from the temporal asymmetry induced by the short-pulse laser field.

Following a similar analysis, Fig. 4 presents the results for $\theta = 15^{\circ}$ and $30^{\circ}$.
At $\theta = 15^{\circ}$ (Fig. 4(a)), the long and short orbits of branches $\textrm{B}_1$ and $\textrm{B}_4$ converge near H20, while those of $\textrm{B}_2$ and $\textrm{B}_3$ converge at H30 and H40, respectively.
These features are also consistent with the enhanced harmonics marked by the star, rhomboid, and triangle at $\theta = 15^{\circ}$ in Fig. 2 (a).
Similarly, Figs. 4(c) and 4(d) show that at $\theta = 30^{\circ}$, convergence occurs for branches $\textrm{B}_2$ and $\textrm{B}_3$ at H33 and H42, in qualitative agreement with the enhanced harmonics observed in Fig. 2(a).
Moreover, at $\theta = 30^{\circ}$, the trajectories of $\textrm{B}_1$ and $\textrm{B}_4$ almost completely overlap from H10 to H20, which explains the enhanced H16 marked by the star in Fig. 2(a).

\begin{figure}[t]
\begin{center}
\includegraphics[width=8.5cm,height=7cm]{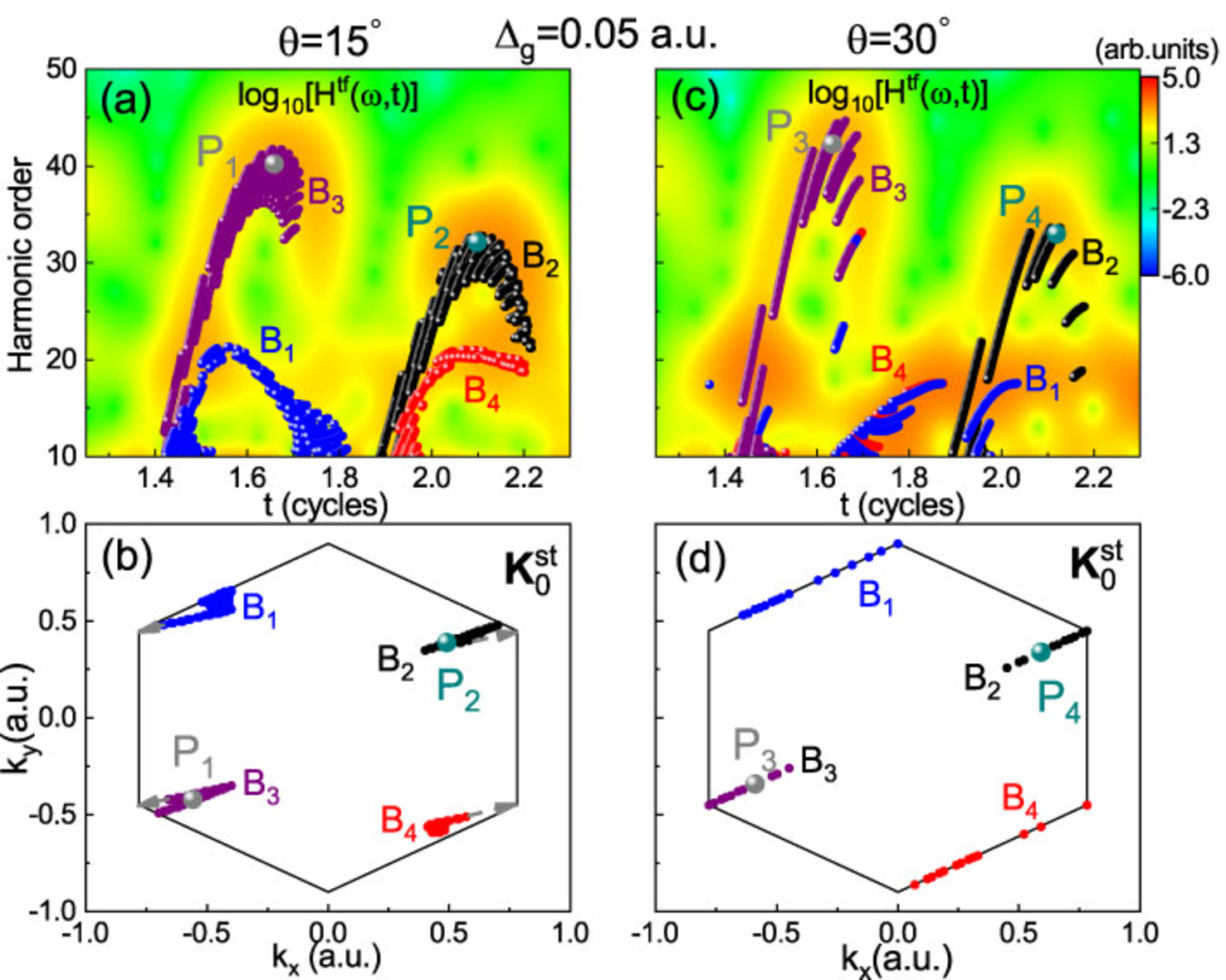}
\caption{Same as Fig. 3, but for orientation angles of $\theta = 15^\circ$ and $\theta = 30^\circ$.
The points $\textrm{P}_1$, $\textrm{P}_2$, $\textrm{P}_3$, and $\textrm{P}_4$ mark four representative electron trajectories.
}
\label{fig:graph1}
\end{center}
\end{figure}

\subsection{Orientation dependence of the ellipticity of the enhancement harmonic}

Next, we select specific orientation angles of $\theta = 15^{\circ}$ and $\theta = 30^{\circ}$ to analyze the orientation dependence of the ellipticity of the enhanced harmonics marked by small triangles and rhomboids in Fig. 2(b).
At $\theta = 15^{\circ}$, the ellipticities of H32 and H40 are relatively low, especially for H40, whose ellipticity is nearly zero.
In contrast, at $\theta = 30^{\circ}$, the ellipticities of H33 and H42 are significantly higher.

\begin{figure}[t]
\begin{center}
\includegraphics[width=8.5cm,height=9cm]{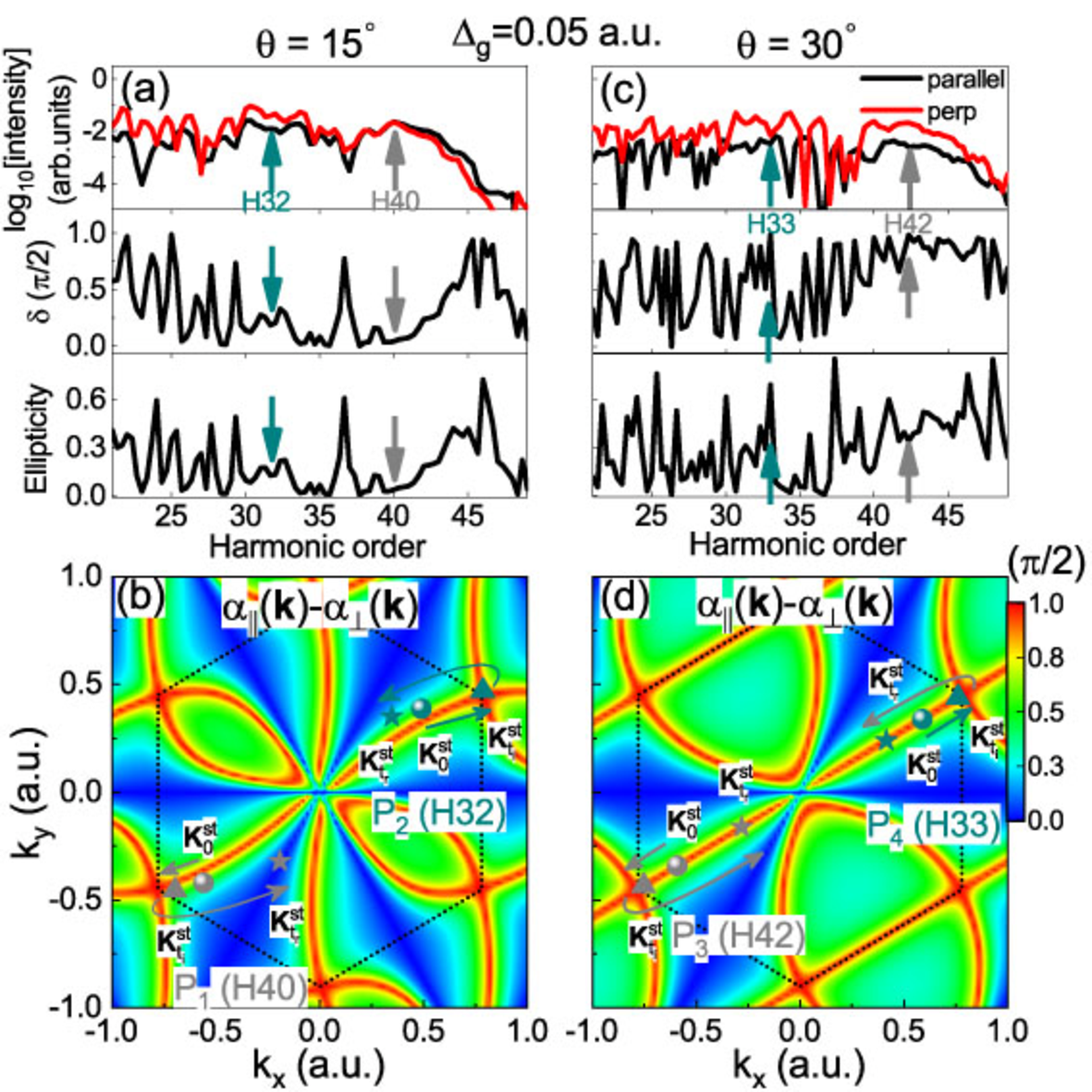}
\caption{(a) Parallel and perpendicular harmonic spectra (upper panel), phase difference between the two spectra (middle panel), and ellipticity (lower panel), calculated for gapped graphene with $\Delta g = 0.05$ a.u. at $\theta = 15^{\circ}$.
The gray and cyan arrows indicate the fortieth (H40) and thirty-second (H32) harmonics.
(b) Phase difference $\alpha_{\parallel}(\textbf{k}) - \alpha_{\perp}(\textbf{k})$ for $\theta = 15^{\circ}$.
For trajectories $\textrm{P}_1$ and $\textrm{P}_2$, the corresponding saddle-point momenta $\textbf{\textrm{K}}_{0}^{\textrm{st}}$, $\textbf{\textrm{K}}_{t_i}^{\textrm{st}}$, and $\textbf{\textrm{K}}_{t_r}^{\textrm{st}}$ are marked.
(c), (d) Same as (a) and (b), respectively, but for an orientation angle of $\theta = 30^{\circ}$.
}
\label{fig:graph1}
\end{center}
\end{figure}

In Fig. 5(a), we show the parallel and perpendicular harmonic spectra, the phase difference between them, and the resulting harmonic ellipticity at $\theta = 15^{\circ}$.
For H40, although the yields of the parallel and perpendicular components are nearly identical, the phase difference is close to zero.
As a result, the ellipticity of H40 remains very low.
Moreover, it can be observed that the trends of the phase difference and the ellipticity with respect to harmonic order are approximately consistent, implying that the phase difference plays a crucial role in determining the ellipticity.

According to the electron-hole recombination trajectory model, the phase difference depends on $\alpha_{\parallel}(\mathbf{K}^{\textrm{st}}_{t_r})-\alpha_{\perp}(\mathbf{K}^{\textrm{st}}_{t_r})$ as described by Eq. (9).
Therefore, in Fig. 5(b), we plot $\alpha_{\parallel}(\mathbf{k})-\alpha_{\perp}(\mathbf{k})$ and the saddle-point solutions of two trajectories $\textrm{P}_1$ and $\textrm{P}_2$, which correspond to H40 and H32, respectively,
and are also marked in Figs. 4(a) and 4(b).
As indicated by the gray star, for trajectory $\textrm{P}_1$, the phase difference predicted by our model is $0.06$ radians, in good agreement with the numerical result of $0.07$ radians labeled by the gray arrow in the middle panel of Fig. 5(a).
For trajectory $\textrm{P}_2$, the model predicts a phase difference of $0.45$ radians (marked by the cyan star), which is close to the numerical result of 0.33 radians indicated by the cyan arrow.

For $\theta = 30^{\circ}$, we focus on H33 and H42 to analyze the generation mechanism of relatively high ellipticity.
In contrast to the case of $\theta = 15^{\circ}$, the phase differences for both H33 and H42 are close to $\pi/2$, which plays a dominant role in the high ellipticity.
As shown in Fig. 5(d), our model also predicts that the phase differences
for trajectories $\textrm{P}_{1}$ and $\textrm{P}_{2}$ are close to $\pi/2$, as marked by the gray and cyan stars, respectively.
(Note that for the enhancement harmonic marked by small stars in Fig. 2(a), because their ellipticity is influenced by contributions from different branches $\textrm{B}_1$ and $\textrm{B}_4$, the generation mechanism is more complex.
Therefore, these cases are not discussed in detail here.)

\section{Synthesis of isolated elliptically polarized attosecond pulses}
\label{s4}

\begin{figure}[t]
\begin{center}
\includegraphics[width=8.5cm,height=11cm]{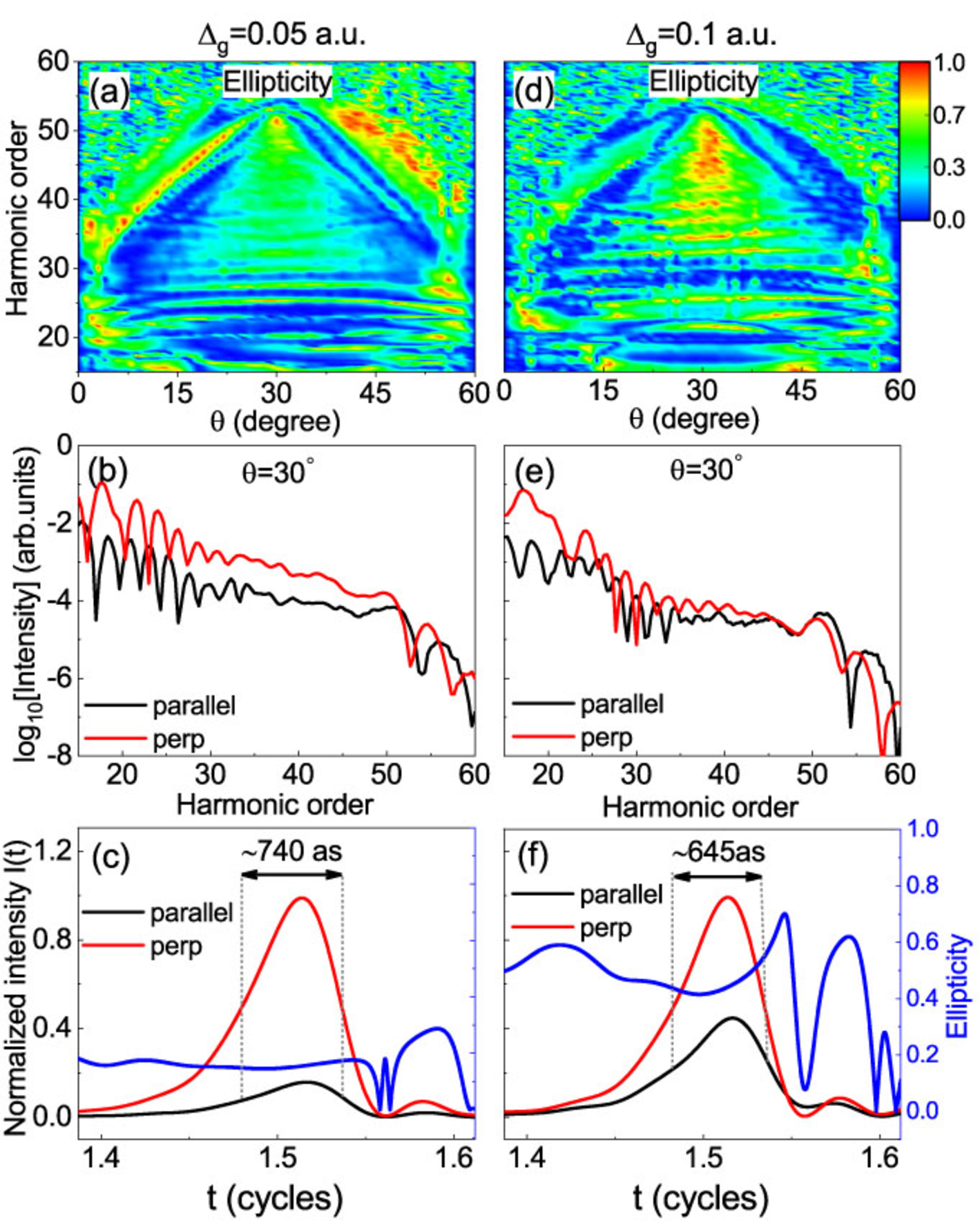}
\caption{
(a) Harmonic ellipticity as a function of the orientation angle, calculated using the laser field defined in Eq. (10) for gapped graphene with $\Delta g = 0.05$ a.u.
(b) Parallel and perpendicular harmonic spectra at an orientation angle of $\theta = 30^{\circ}$.
(c) Attosecond pulses synthesized from parallel and perpendicular harmonics shown in panel (b).
The blue curves indicate the ellipticity of the attosecond pulse, while
the dotted lines mark the half-maximum position of the two pulses.
(d)-(f) Same as (a)-(c), respectively, but for $\Delta g = 0.1$ a.u.
}
\label{fig:graph1}
\end{center}
\end{figure}

Since the laser is linearly polarized, the electrons oscillate along the polarization direction in reciprocal space under its influence.
By examining Figs. 4 (d) and 5 (d), one can conclude that for $\theta = 30^{\circ}$, the phase difference associated with the trajectories on branch $\textrm{B}_3$ remains consistently close to $\pi/2$.
Therefore, if the laser field can be engineered to selectively amplify the spectral contributions from the branch $\textrm{B}_3$ in Fig. 4(c), while suppressing those from other branches, one may obtain IEAPs.
Here, we design a two-color (fundamental plus second-harmonic) field scheme described by

\vspace{-0.4cm}
\begin{align}
\mathbf{F}^{\prime}(t) =F_0 f(t) [ \cos(\omega_0 t) - 0.7 \cos(2 \omega_0 t) ] \textit{\textbf{e}}.
\end{align}

Figure 6(a) presents the harmonic ellipticity simulated using the laser described by Eq. (10).
At $\theta = 30^{\circ}$, the harmonic ellipticity remains around $0.5$ from H30 to H53.
As shown in Fig. 6(b), within this harmonic range, the intensity difference between the perpendicular and parallel components ranges from approximately $0.4$ to $1.2$ orders of magnitude, which primarily limits the achievable ellipticity.

Figure 6(c) shows the parallel and perpendicular attosecond pulses, with a full width at half maximum (FWHM) of approximately $740$ as, synthesized using the intensity $I_{\mu}(t) = |\epsilon_{\mu}(t)|^2$ and phase $\varphi_{\mu}(t) = \textrm{arg}[\epsilon_{\mu}(t)]$.
The synthesized pulse $\epsilon_{\mu}(t)$ in the plateau region is given by
$\epsilon_{\mu}(t) = \sum_{q=q_{\textrm{L}}}^{q_{\textrm{H}}} E_{\mu}^q \exp\left[i(q\omega_0 t + \phi_{\mu}^q)\right]$, in which $q_{\textrm{L}} = 30$, $q_{\textrm{H}} = 53$, $E_{\mu}^q$ and $\phi_{\mu}^q$ denote the amplitude and phase of the $q^{\textrm{th}}$ harmonic in $\mu$ direction.
It can be observed that the amplitude of the parallel component is much lower than that of the perpendicular component, resulting in an ellipticity of approximately $0.2$ as illustrated by the blue line in Fig. 6(c).

To further improve the ellipticity of the pulse, we perform additional simulations using gapped graphene with an energy gap of $\Delta_g = 0.1$ a.u., irradiated by the same two-color field described in Eq. (10).
As shown in Fig. 6(d), the ellipticity of the harmonics from the $30^{\textrm{th}}$ to the $53^{\textrm{rd}}$ order is significantly enhanced, ranging from $0.5$ to $0.9$.
At $\theta = 30^{\circ}$, the yield of perpendicular harmonics becomes comparable to that of parallel harmonics, as illustrated in Fig. 6(e).
In Fig. 6(f), the synthesized attosecond pulses, with a FWHM of approximately 645 as, exhibit a markedly improved amplitude balance between the parallel and perpendicular components compared to Fig. 6(c), resulting in increased ellipticity close to $0.5$.

\section{Conclusion}
\label{s5}

In summary, we have investigated the underlying mechanisms of enhanced harmonics and their associated ellipticity in gapped graphene irradiated by a short-pulse femtosecond laser at different orientation angles.
By combining a recombination trajectory model with time-frequency analysis, we reveal that the harmonic enhancement originates from the convergence of long and short orbits, known as the caustic effect.
The orientation dependence of the enhanced harmonics is attributed to the inequivalence of the $\textrm{K}$ points involved in the electron ionization process.
We further analyze the ellipticity of the enhanced harmonics at selected angles, and demonstrate that it primarily depends on the phase difference between the parallel and perpendicular components, which can be accurately predicted by our recombination trajectory model.
Based on these insights, we propose a strategy to generate IEAPs.
By employing a two-color field scheme, we successfully synthesize elliptically polarized pulses with FWHM of approximately 740 and 645 attoseconds in gapped graphene with $\Delta_g = 0.05$ a.u. and $0.1$ a.u., respectively.
This work may shed light on the generation of IEAPs in gapped graphene or TMDs.

\section*{ACKNOWLEDGMENTS}

This work is supported by the NSAF (Grants No. 12404394, and No. 12347165),
Hebei Province Optoelectronic Information Materials Laboratory Performance Subsidy Fund Project (No. 22567634H), and the High-Performance Computing Center of Hebei University.

\end{document}